# ROLE OF POINT DEFECTS ON THE PROPERTIES OF MANGANITES

**Lorenzo Malavasi*[a], Maria Cristina Mozzati[b], Paolo Ghigna[a], Gaetano Chiodelli[a], Carlo B. Azzoni[b] and Giorgio Flor[a]**

[a] *Dipartimento di Chimica Fisica "M. Rolla", INSTM, IENI/CNR Unità di Pavia of Università di Pavia, V.le Taramelli 16, I-27100, Pavia, Italy.*
**E-mail: malavasi@chifis.unipv.it*
[b]*INFM, Unità di Pavia and Dipartimento di Fisica "A. Volta", Università di Pavia, Via Bassi 6, I-27100, Pavia, Italy.*

## Abstract

$La_{1-x}Ca_xMnO_{3+\delta}$ and $La_{1-x}Na_xMnO_{3+\delta}$ samples with well defined cation molecularity and oxygen contents are analyzed with XRPD, electrical conductivity, electron paramagnetic resonance and static magnetization measurements. Point defects are introduced in the structure by: i) aliovalent cation doping (substitutional defects), ii) oxygen over-stoichiometry (cation vacancies) and iii) oxygen under-stoichiometry (oxygen vacancies). The cation doping mainly influences the material by affecting the tolerance factor and the oxidation of the Mn ions; the cation vacancies affect the magnetic properties by interrupting the interaction paths between $Mn^{4+}$-$Mn^{3+}$ ions whereas the oxygen vacancies have a stronger influence on the structural and electrical properties and act on the magnetic properties by the overall decrease on the $Mn^{4+}$ amount and by the creation of sample regions with different magnetic features. We can state that the basic magnetic and electrical behaviors of the Ca- and Na-doped lanthanum manganite compounds are driven from the $Mn^{4+}$-$Mn^{3+}$ ratio and all the peculiar behaviors are indeed caused by defects, concentration gradient and, more generally, by lattice disorder.

# Introduction

Rare Earth manganite perovskites of general formula $Ln_{1-x}A_xMnO_3$ (Ln = La, Pr; A = Ca, Sr) are, at present, among the most studied materials due to their relevant properties such as the colossal magneto-resistance (CMR) effect and charge and spin order. In fact, the renewed interest about these oxides, first studied experimentally by Jonker and Van Santen and theoretically by Zener and Goodenough in the '50s [1-3], came out from the observation of a huge reduction of the electrical resistivity of divalent-doped $LaMnO_3$ [4] when applying a magnetic field. Later it was pointed out the role played by the different oxidation states, +3 and +4, of the manganese ions in these oxides which seemed to be the main variable in tuning the physical properties of manganites: in $LaMnO_{3+\delta}$ and $La_{1-x}Ca_xMnO_3$ systems, for example, a $Mn^{4+}$ content around ~30% was found to be the optimal value in order to achieve the maximum CMR effect and also the highest Curie temperature ($T_c$) for the paramagnetic to ferromagnetic transition (P-F) [5,6].

In the magnetic ordered state the electron hopping process is favored by the spin arrangement while for temperatures above $T_c$ the conduction mechanism is semiconducting-like. This behavior was closely related to the double-exchange (DE) mechanism, first proposed by Zener [3], which involves the transfer of $e_g$ electrons between neighboring Mn ions when these are ferromagnetically coupled. The DE mechanism alone does not allow a satisfactory description of the resistivity trend in these systems and a more complete picture takes into account the electron-phonon coupling associated with the Jahn-Teller (J-T) effect [7,8].

Usually, the $Mn^{4+}$ amount is tuned by aliovalent cation doping on the A site of the perovskite structure. Another way to change the $Mn^{4+}$-$Mn^{3+}$ ratio is the variation of the oxygen content ($\delta$) by controlling temperature ($T$) and oxygen partial pressure ($P(O_2)$) during annealing steps. Anyway, the effect of the $\delta$-variation has a different nature with respect to the cation doping since it acts not only as a holes or electrons source but also by creating point defects within the lattice [9]. These point defects may play a fundamental role in affecting the physical properties of manganites, *i.e.* their crystallographic structure, the Curie temperature, the electric behavior and the magnetoresistive response.

The main sources of point defects on manganites are the aliovalent cation doping and the oxygen non-stoichiometry. Concerning the first, its direct effect is on the structural properties of the manganite by the variation of the tolerance factor (see later in the text). This in turn affects the electronic structure of the material. Secondly, the extent of the cation doping determines, as said before, the oxidation state of the Mn array. Our attention will be played in the following particularly on the defects associated with the oxygen under- or over-stoichiometry since the role played by the cation doping is better assessed.

With respect to the role played by the oxygen content on the physical properties we can say that manganites share several properties with the high temperature superconductors (HTSC) of general formula $REBa_2Cu_3O_y$. Much of the research actually carried out on manganites exploits the experience accumulated on cuprate oxides for which the fundamental role of oxygen content control was clearly assessed [10-18]. Up to now this features have not yet considered with the correct relevance also in the field of magnetoresistive manganites where an appreciable variation of the oxygen content is present.

The aim of this short review is to present the results of the experimental investigations carried out by our research group on two manganites systems, $La_{1-x}Ca_xMnO_{3+\delta}$ and $La_{1-x}Na_xMnO_{3+\delta}$, with divalent and monovalent doping respectively, in order to unveil the role of oxygen non-stoichiometry on their physical properties.

# Results and Discussion

*1. Defect Chemistry of Manganite Perovskite*

The magnetoresistive manganites of general formula $LnMnO_{3+\delta}$ present a perovskite structure. The ideal $ABO_3$ perovskite is cubic and it is constituted of a closed packed array of oxygen ions in which the cations are placed. The larger A ion, usually a Rare Earth (RE), has a N = 12 coordination number while the B cation (manganese) is octahedrally coordinated (N = 6). Actually, the most common manganites, where Ln = La or Pr, are not cubic but somehow distorted towards less symmetric structures. This results from the mismatch of the equilibrium (Mn-O) and (A-O) bond lengths and can be rationalized by considering the *tolerance factor* (*t*) which is defined as:

$$t = \frac{(A-O)}{\sqrt{2}(Mn-O)} \tag{1}$$

In the manganite perovskite considered in the following, *i.e.* where the A ion is the lanthanum, the structure usually adjusts to $t < 1$ by a cooperative rotation of the $MnO_6$ octahedra thus giving origin to Mn-O-Mn bond angles lower than 180° and to a rhombohedral (if the rotation is around the [111] axis) or orthorhombic (if the rotation is about a cubic [110] axis) structure. The possibility, for the perovskite structure, of this structural distortions allows rich and extensive cation replacements on both the A and B site of the unit cell; the structure is also stable with respect to high level of oxygen and cation vacancies. Up to now the most interesting and studied kind of cation substitutions regard the ones on the A site; in fact, it was on strontium and calcium doped $LaMnO_{3+\delta}$ that the first evidences of Colossal Magnetoresistivity (CMR) were found [4,19,20]. Anyway, also the undoped $LaMnO_{3+\delta}$ presents the CMR effect.

The meaning of this kind of replacement resides on the oxidation of the Mn-ions array caused by the aliovalent substitution. For the undoped member the oxidation of the manganese comes from the variation of the oxygen content. Of course, also for the cation-doped samples the oxidation of manganese may be caused by the $\delta$-variation so giving origin to a contribution to the $Mn^{4+}$ formation by cation replacement and by oxygen content variation.

The defect chemistry of CMR manganites has been first considered for the undoped $LaMnO_{3+\delta}$ by Van Roosmalen *et al.* [21,22]. They determined that pure lanthanum manganese oxide always presents an oxygen over-stoichiometry compensated by cation vacancies, equally distributed between lanthanum and manganese sites according to the following equilibrium:

$$\tfrac{1}{2}O_2 \rightleftarrows O_O^x + \tfrac{1}{3}V_{La}''' + \tfrac{1}{3}V_{Mn}''' + 2h^{\bullet} \tag{2}$$

Neutron diffraction experiments confirmed this model, even though there is still some debate about the relative amount of the lanthanum and manganese vacancies [23,24]. It is to remark that the closed-packed arrangement of the perovskite structure does not allow the formation of interstitial oxygen. Also computational simulation clearly showed that the main defects which can be expected on lanthanum manganites are cation vacancies and not interstitials ions [25]. So, according also to these simulations, the oxygen over-stoichiometry will be accommodated by cation vacancies as presented in equation (2).

For this reason we should note that the presence of oxygen over-stoichiometry is not properly accounted for in the formalism $LaMnO_{3+\delta}$ and the correct formula for these compounds should be: $La_{1-\varepsilon}Mn_{1-\varepsilon}O_3$ where $\varepsilon = \delta/(3+\delta)$. Anyway, this notation, although correct, is less straightforward than the one usually adopted in the current scientific literature regarding manganites, *i.e.* $LaMnO_{3+\delta}$. As a consequence we are going to use this last formalism having in mind, anyway, the correct defect chemistry of oxygen over-stoichiometric materials.

The holes created by the oxygen excess will be responsible for the oxidation of the $Mn^{3+}$ ions:

$$Mn_{Mn} + h^{\bullet} \leftrightarrows Mn_{Mn}^{\bullet} \tag{3}$$

These charge carriers may be localized-$e$ or itinerant-$\sigma^*$ electrons according to the relative magnitude of the electron-phonon interactions and electron interactions with the oxygen atom vibration, which trap the charge carriers, and the frequency of the electron transfer [26]. In other terms, for a given energetic magnitude of the bandwidth of the conduction band made of the Mn-$e_g$ and O-$p$ orbitals ($W_\sigma$) it can be found that the polaronic (localized) regime corresponds to the situation in which $W_\sigma < \hbar\,\omega_R$, where $\omega_R$ is the frequency of the lattice vibration (phonon) which traps the charge carriers, while the itinerant regime is found for $W_\sigma > \hbar\,\omega_R$ [27].

According to the above equilibria it is clear that an oxygen under-stoichiometry or a reduction of the $\delta$-value should cause a reduction of manganese 4+ by increasing the $Mn^{3+}/Mn^{4+}$ ratio.

The presence of a dopant on the lanthanum site with a lower oxidation number, as mentioned before, is responsible for the oxidation of the manganese ions. For the $Ca^{2+}$ doping the pertinent chemical equilibrium is:

$$2CaO + 2La_{La} + 6O_O \rightleftarrows 2Ca'_{La} + La_2O_3 + V_O^{\bullet\bullet} + 5O_O \tag{4}$$

the oxygen vacancies annihilation by equilibrium with the external atmosphere can be expressed as:

$$2V_O^{\bullet\bullet} + O_2 \rightleftarrows 2O_O + 4h^{\bullet} \tag{5}$$

So, for the general composition $La_{1-x}Ca_xMnO_{3+\delta}$ and taking into account all the equilibria presented above the number of $Mn^{4+}$ is then:

$$[Mn^{\bullet}_{Mn}] = (x+2\delta) \tag{6}$$

Moreover, as the cation-doping introduces steric effects due to the ionic radii mismatch, which have influence also on the electronic structure of the material via the tolerance factor [26, 27], the oxygen content variation introduces point defects which may be trapping centers and lattice distortion for the charge carriers.

In the case of a monovalent dopant, such as sodium, the corresponding quasi-chemical relation for the substitutional defect is:

$$Na_2O + 2La_{La} + 6O_O \rightleftarrows 2Na''_{La} + La_2O_3 + 2V_O^{\bullet\bullet} + 4O_O \tag{7}$$

and the overall amount of $Mn^{4+}$ is then:

$$[Mn^{\bullet}_{Mn}] = 2(x+\delta) \tag{8}$$

So, for the sodium doping the amount of holes introduced, for the same level of cation substitution ($x$), is twice with respect to the one produced by the calcium doping. Moreover, the tolerance factor for the Sodium-doped samples is greater relative to the Ca-doped samples, *i.e.* closer to the ideal value of 1.

In the following the role played by the oxygen non-stoichiometry, cation vacancies, and the cation doping for the $La_{1-x}Ca_xMnO_{3+\delta}$ and $La_{1-x}Na_xMnO_{3+\delta}$ systems will be presented. The study of the combined effect of substitutional defects and oxygen content variation on the physical properties of manganites has scarcely carried out in the current literature. On the opposite, the effect of the $\delta$-variation on the pure the $LaMnO_{3+\delta}$ has been object of a grater amount of work [27-29].

## 2. $La_{1-x}Ca_xMnO_{3+\delta}$

Figure 1 reports the X-ray diffraction patterns for the six samples considered [30]. The lattice constants are reported in Table 1. For the samples with $x$ = 0.10, $\delta$ = -0.014 and $\delta$ = 0.0 the distortion is orthorhombic (space group *Pbnm*) and of *O'*-type since $c/a<\sqrt{2}$. In this phase the Jahn-Teller distortion, cooperative and static, is correlated with the $e_g$ $(x^2-y^2)$ orbital ordering, which expands the unit cell in the *a-b* plane [31]. For the sample with $x$ = 0.1 and $\delta$ = 0.054 and for the samples with $x$ = 0.3, $\delta$ = -0.01 and $\delta$ = –0.025 the crystal structure is the *O*-type orthorhombic one, resulting from the $e_g$ $(3z^2\text{-}r^2)$ orbital ordering, expanding the unit cell in the *c* direction ($c/a>\sqrt{2}$). For the sample with $x = 0.3$ and $\delta = 0$ the structure is rhombohedral ($R\bar{3}c$).

From the data in Table 1 it is possible to deduce that when both Ca doping and oxygen content increase the cell volume decreases. This effect is directly related to the progressive reduction (increase) of $Mn^{3+}$ ($Mn^{4+}$) as the oxygen content and Ca doping increase and on the ionic radii differences between $Mn^{3+}$ and $Mn^{4+}$ ions: 0.58 Å and 0.53 Å, respectively. Our data agree with the literature ones [32], for which the crossover between the *O*' and the *O*-type structures is around 10% of $Mn^{4+}$ ions.

Figure 2 presents the $\rho$ *vs.* $T$ plots for the $x = 0.1$ samples. For the $La_{0.9}Ca_{0.1}MnO_3$ (a) and $La_{0.9}Ca_{0.1}MnO_{2.986}$ (b) samples the transport behavior is semiconducting-like in all the *T*-range explored. The asterisks in the graph mark shoulders falling at temperatures, reported in brackets in Table 1, corresponding to the magnetic transition temperatures ($T_c$), as will be shown later.

For the $La_{0.9}Ca_{0.1}MnO_{3.054}$ sample, (c), the resistivity curve shows a transition from a semiconducting to a metallic-like regime (S-M). The transition temperature ($T_\rho$), taken at the maximum of the peak, is about 210 K. The transition appears quite broad and a second peak may be identified at about 190 K. Moreover, the resistivity at the lower investigated temperature is higher compared to the one at room temperature.

The fitting of the semiconducting-like regime, by considering the following adiabatic small-polaron thermally activated hopping process:

$$\rho = \rho_0 T \exp(E_a / k_B T) \qquad (9)$$

allows us an estimation of the activation energies ($E_a$). The $E_a$ values for the samples with $x = 0.1$, reported in Table 2, increase with the decreasing of the oxygen content. The very easier hopping for the sample with $\delta = 0.054$ clearly reflects the less distorted crystal structure (*O*-type) with respect to the other two samples (*O'*-type).

The metallic-like regime was followed by interpolating the curve for $T<0.85T_\rho$ (to assure the completion of the S-M transition) with the model currently proposed [6,33]

$$\rho = \rho_0 + \rho_2 T^2 + \rho_{4.5} T^{4.5} \qquad (10)$$

where $\rho_0$ represents the *T*-independent contribution to $\rho$, due mainly to grain boundaries and point-defects; the term $\rho_2 T^2$ takes into account the electron-electron scattering processes while the $\rho_{4.5} T^{4.5}$ term considers the electron-magnon interaction. The results of the fit for the sample with $\delta = 0.054$ are reported in Table 2.

The plots of $\rho$ *vs.* $T$ for the samples with $x = 0.3$ are reported in Figure 3. All the samples present a S-M transition with $T_\rho$ strongly dependent on the $\delta$-values. Also the width of the transitions is $\delta$-dependent, increasing with reducing the oxygen content.

In detail, for $\delta = 0$ $T_\rho = 271$ K, in good agreement with the available published results [6]; for $\delta = -0.01$, a sharp S-M transition occurs at 261 K and a second one, broader, at 239 K; for $\delta = -0.025$ $T_\rho$ is reduced to 229 K, the transition is very broad and a shoulder may be detected at about 257 K. $T_\rho$-values and shoulder temperatures (in brackets) are reported in Table 1.

The $E_a$ values for $x = 0.3$ samples, as deduced by (9), are reported in Table 2: it is clear that they increase as the oxygen content decreases.

For the same samples Table 2 reports the results of the fit for the metallic regime according to (10). The oxygen under-stoichiometry, *i.e.* the presence of oxygen vacancies, deeply affects the transport properties resulting in both $Mn^{4+}$ reduction and point defects creation within the structure. Indeed, the $\rho_0$ value increases by reducing first the Ca doping and second the oxygen content, the lowest value being achieved for $La_{0.7}Ca_{0.3}MnO_3$. About the other resistivity parameters, we observe that the $\rho_2$ value is extremely low for the $x = 0.1$ sample while the opposite holds for the $\rho_{4.5}$ value. For the $x = 0.3$ samples the two values increase by reducing the oxygen content, being the $\rho_{4.5}$ more sensitive to this effect. A possible explanation of the different dominant scattering process may be found in the effect of Ca doping on the band structure. Indeed, the increase of Ca-doping causes an increase of the tolerance factor (*t*) and consequently an increase of the intrinsic $Mn^{4+}$ amount, giving rise, according to [31], to an increase of the band width formed by O-*2p* and Mn-$e_g$ orbitals. Moreover, as will be shown later, the sample with $x$ = 0.1 is characterized by smaller magnetic domains with respect to the $x$ = 0.3 samples. All these facts point to an increase of intra-band scattering phenomena between charge carriers for the $x$ = 0.1 sample with respect to the $x$ = 0.3 ones.

Figure 4 shows the static molar magnetization ($M_{mol}$) *vs.* $T$ data acquired at 100 Oe for the $x$ = 0.1 samples. A progressive decrease of the Curie temperature ($T_c$) for the P-F transition is observed with the reduction of the oxygen content. The $T_c$ values are listed in Table 1. Also the width of the magnetic transition increases in the same direction, as evidenced by the field cooled (FC) curves. Concerning the zero field cooling (ZFC) curves, a trend from a situation of large ferromagnetic clusters ($\delta$ = 0.054) towards a spin-glass-type behavior ($\delta$ = -0.014) can be observed. For the sample with the lowest $Mn^{4+}$ content and also with a relatively large amount of anionic vacancies, the ZFC magnetization nearly vanishes by lowering the temperature indicating a large amount of randomly oriented magnetic clusters.

The molar magnetization at 100 Oe of the samples with $x$ = 0.3 is shown in Figure 5. A $T_c$ value of 270 K pertains to the sample with $\delta$ = 0 while for the other two samples $T_c$ is about 258 K (Table 1). For all these samples the ZFC behavior suggests the presence of long range magnetic structure. The difference between FC and ZFC is greater for the sample with the highest oxygen vacancies amount, *i.e.* for $\delta$ = -0.025.

*3. $La_{1-x}Na_xMnO_{3+\delta}$*

Figure 6 reports the X-ray powder diffraction (XRPD) patterns for the samples with $x$ = 0.05 and $\delta$ = 0.078, 0, and -0.01 and for the samples with $x$ = 0.13 and $\delta$ = 0, -0.03, and -0.04. For $x$ = 0.05 the crystal structure is orthorhombic when $\delta$ = 0 and $\delta$ = -0.01 and can be indexed according to the *Pnma* structure (Space Group n. 62) while for $\delta$ = 0.078 the crystal structure is more symmetric belonging to the rhombohedral $R\bar{3}c$ space group. Moreover, for $La_{0.95}Na_{0.05}MnO_{2.99}$ the lattice parameters ratio ($c/a < \sqrt{2}$) suggests the presence of a static cooperative Jahn-Teller deformation of the octahedra and an orthorhombic *O'* structure. For the sample $La_{0.95}Na_{0.05}MnO_3$ $c/a \sim \sqrt{2}$, thus suggesting the evolution towards a dynamic J-T deformation by increasing the oxygen content. For $x$ = 0.13 all the samples are rhombohedral. The correspondent lattice constants and cell volumes are reported in Table 1 for all the samples.

By looking at the cell volumes of Table 1 it is possible to note that by increasing the sodium content, for the same oxygen content, the cell dimension reduces and by increasing the oxygen content, at fixed $x$, the cell volume reduces too. Since the ionic radii of $La^{3+}$ (1.36 Å) and $Na^{+1}$ (1.39 Å), for the same coordination, are very close [34], this reflects the progressive increase of the $Mn^{4+}/Mn^{3+}$ ratio by increasing $\delta$ and/or $x$ being the ionic radii 0.645 Å and 0.53 Å for $Mn^{3+}$ and $Mn^{4+}$, respectively [34]. This allowed us to concentrate mainly on the role of the oxygen non-stoichiometry effect and less to the effect of the average ionic radius of the A site responsible of the Mn-O bandwidth [26] through the tolerance factor *t*. The evolution from an orbitally ordered orthorhombic structure to a rhombohedral one, as the cation doping or/and oxygen over-stoichiometry increases, reflects the progressive decrease of J-T $Mn^{3+}$ ions, the subsequent reduction of (Mn-O) bond length and the approach to 180° of Mn-O-Mn bond angle which is the optimal condition for the charge carrier hopping.

Figure 7 presents the plot of the electrical resistivity ($\rho$) as a function of temperature for the $La_{0.95}Na_{0.05}MnO_{3+\delta}$ samples. Measurements were performed in a reduced temperature range because of the high resistivity values. As can be inferred, all the samples present an activated transport. Only for the highest oxygen content ($\delta$ = 0.078) a transition to a metallic-like conductivity is detectable at $T_\rho$ ~120 K. In spite of the large difference in the $Mn^{4+}$ content, the samples with $\delta$ = 0.078 and $\delta$ = 0 display similar $\rho$-values, lower than that of the under-stoichiometric sample.

Figure 8 reports the $\rho$ *vs.* $T$ behavior for the samples with $x$ = 0.13. The samples with $\delta$ = 0 and $\delta$ = –0.03 show a clear transition from a semiconducting-like (activated) regime to a metallic-like one (S-M) at $T_\rho$ ~303 K and $T_\rho$ ~260 K respectively (as determined from the derivative of the $\rho(T)$ curve). The third sample ($\delta$ = -0.04) displays a transition as well, even if broader than in the former case, between two activate regimes. Let's note that this sample is the one with the highest oxygen under-stoichiometry.

The semiconducting-like regime, dominated by the charge carriers trapping phenomena related to lattice distortions and/or defects, has been analyzed be means of an adiabatic small-polaron model, according to equation (9).

For the $La_{0.95}Na_{0.05}MnO_{3+\delta}$ samples the $E_a$ values range around 200 meV. It is interesting to note, anyhow, that the sample with the lowest $Mn^{4+}$ content and oxygen vacancies shows the highest $E_a$ value. For the $La_{0.87}Na_{0.13}MnO_{3+\delta}$ samples the activation energies are smaller, as a consequence of the less distorted crystal structure which enables an easier polaron dynamic. Moreover a clear trend with the oxygen content may be now emphasized.

The metallic regime was fitted by using the empirical relation presented above (10). The values of $E_a$ and the parameter for the metallic zones are reported in Table 2.

Figure 9 and 10 present the static magnetization ($M$) behavior *vs.* $T$ for the $La_{0.95}Na_{0.05}MnO_{3+\delta}$ and $La_{0.87}Na_{0.13}MnO_{3+\delta}$ series, respectively. FC and ZFC measurements were performed at 100 Oe. For all the samples a P-F transition occurs. The Curie temperatures ($T_c$), taken at the inflection points in the FC curves, are reported in Table 1.

In detail, for the $x$ = 0.05 samples the maximum $T_c$ value is achieved for the sample with $\delta$ = 0 (152 K) while for the other two samples the $T_c$ are very close each other. Thus, an easy trend of $T_c$ *vs.* $\delta$, and consequently *vs.* the $Mn^{4+}$ content, does not hold: the induced defects strongly influence the $T_c$ values, nearly irrespectively of the $Mn^{4+}$ amount. The shape of the ZFC curves (Figure 9) suggests the presence of magnetic domains, randomly oriented when cooling without an applied magnetic field. This is particularly evident for the under-stoichiometric sample ($\delta$ = -0.01). Finally, let's note the relatively broadness of the magnetic transitions for these samples.

The $La_{0.87}Na_{0.13}MnO_{3+\delta}$ samples show sharper magnetic transitions and reduced differences between FC and ZFC curves (Figure 10) compared to the $x$ = 0.05 ones. The highest $T_c$ pertains to the sample with the correct oxygen stoichiometry, *i.e.* defects free, which is also the one with the sharpest P-F transition and the smallest difference between FC and ZFC curves. The other two samples have approximately the same $T_c$ even if the sample with the greater oxygen vacancies content, $\delta$ = -0.04, displays a greater difference between FC and ZFC curves.

EPR measurements can aid to deepen the comprehension of the magnetic behavior, particularly when approaching the magnetic transition. In lanthanum manganites the EPR signal may arise from $Mn^{4+}$ single ions or from $Mn^{3+}$- $Mn^{4+}$ Zener pairs [35]: in both cases, in the paramagnetic phase, a signal centered at $g \cong 2$ is expected.

A common behavior has been found for the EPR signals of all $La_{1-x}Na_xMnO_{3+\delta}$ samples. A selected example is shown in Figure 11 where the EPR signal *vs.* $T$ of the sample with $x$ = 0.05 and $\delta$ = 0 is reported. Shortly, for temperatures lower than a $T_L$ value only one signal is observed with $g_{eff}$ > 2, thus evidencing the presence of internal magnetic fields resulting from ferromagnetic ordering in the whole sample; for temperatures higher than a $T_H$ value only one signal with $g \cong 2$ is present, suggesting the paramagnetic character of the whole sample; in the intermediate temperature range, $T_H$ - $T_L \equiv \Delta T_c$, at least two distinct signals, with different trends *vs.* $T$, are detectable.

In detail, for the samples with $x$ = 0.05 two magnetic regions (here named A and B) in which the long-range magnetic interactions hold till different temperatures, $T_A$ and $T_B$, are evident. For $\delta$ = 0.078 ($T_L$ < 140 K) $T_A \cong$ 200 K and $T_B \cong$ 330 K, for $\delta$ = 0 ($T_L \cong$ 140 K) $T_A \cong$ 240 K and $T_B \cong$ 330 K, for $\delta$ = -0.01 ($T_L \cong$ 140 K) $T_A \cong$ 330 K and $T_B \cong$ 600 K.

Concerning the samples with $x$ = 0.13, for $\delta$ = 0 two EPR signals are observed at the lowest investigated temperature (293 K) corresponding to two different magnetic regions with $T_A$ < 293 K and $T_B$ = 310 K which rapidly merge: at 310 K the sample shows a pure paramagnetic behavior. Analogous features can be observed for the other samples with $x$ = 0.13 and with $\delta$ = -0.03 ($T_L$ = 260 K, $T_A$ = 270 K, $T_B$ = 300 K) and $\delta$ = -0.04 ($T_L$ = 250 K , $T_A$ = 255 K, $T_B$ = 300 K).

By assuming $T_H = T_B$, we point out that the $\Delta T_c$ range well corresponds to the onset and completion of the magnetic transition and its extent agrees with the broadening of the magnetic transition. For the same $x$-values, the extent of $\Delta T_c$ strongly depends on the oxygen vacancies. Indeed, the narrowest $\Delta T_c$ value corresponds to the samples with the correct oxygen content, *i.e.* $\delta$ = 0, whereas wider transitions are observed for the samples with $\delta < 0$. For $x$ = 0.13 the temperature range of the coexistence of the two EPR signals is highly reduced with respect to the samples with $x$ = 0.05. The multiplicity of EPR signals suggests the presence in the samples of different magnetic regions due to $Mn^{4+}$ concentration gradients, more evident for $x$ = 0.05, for $\delta$ = 0 and also as a results of an inhomogeneous defect distribution for $\delta \neq 0$.

For the $x$ = 0.13 samples, the EPR signals show the main features already observed in analogous CMR calcium manganites with $x$ = 0.33, that is with a nearly equal amount of $Mn^{4+}$ [36-39]. Shortly, by lowering the temperature the EPR linewidth ($\Delta B$) linearly decreases and reaches its minimum value at a temperature, named $T_{\text{onset}} \sim T_B$, greater than $T_c$. For $T < T_{onset}$ $\Delta B$ increases [37,38]. Opposite, for $x$ = 0.05 the $\Delta B$ *vs.* $T$ trend is more complex as a consequence of the extent of the temperature range ($\Delta T_c$) where the two different magnetic regions coexist. Indeed, two minima, well corresponding to $T_A$ and $T_B$, are present. The EPR measurements allow also relating the magnetic behavior in the critical region ($\Delta T_c$) to $\rho(T)$ peaks and slope variations. They can be associated to the $T_A$ and $T_B$ transition temperatures: indeed, we can observe a peak at ~300 K for the sample with $x$ = 0.13 and $\delta$ = 0, a peak at ~270 K for the sample with $x$ = 0.13 and $\delta$ = -0.03 and the slope change at 250 K for the sample with $x$ = 0.13 and $\delta$ = -0.04. So, the complex $\rho(T)$ behavior is also justified from the sample non-homogeneity.

## Conclusion

The whole set of results we gained by this experimental study on the $La_{1-x}Ca_xMnO_{3+\delta}$ and $La_{1-x}Na_xMnO_{3+\delta}$ systems can be better focused by presenting the data as phase diagrams (Figure 12 and 13) in which $T_\rho$ and $T_c$ are plotted against the $Mn^{4+}$ content, as estimated by the oxygen and cation stoichiometry. For the La-Ca-Mn-O samples also data from Ref. 6 are added while some additional points for the Na-doped phase have been taken from Ref. 40.

It is quite interesting to follow the systems evolution *versus* the $Mn^{4+}$ content for both phases and compare their behavior.

The crystal structure evolves from an orbitally ordered, highly distorted, orthorhombic (*O'*) to a less distorted rhombohedral one by increasing the $Mn^{4+}$ content, *i.e.* by decreasing the J-T $Mn^{3+}$ ions, irrespective that this is created by the oxygen over-stoichiometry or by the aliovalent doping. Interestingly, for both systems the lower limit for the existence of this highly distorted structure is around 10% (in $Mn^{4+}$ content). The *O'* orbital ordering found for low $Mn^{4+}$ content has the effect to localize the charge carriers thus giving origin to insulating phase with low $T_c$ and typical ZFC curves suggesting the presence of short-range magnetic order and spin-glass behavior. A clear transition temperature in the ZFC curves can in fact be only observed for the sample $La_{0.9}Ca_{0.1}MnO_{2.986}$ and $La_{0.95}Na_{0.05}MnO_{2.99}$ (see Table 1 for samples details). The ferromagnetic component, which is established between two-manganese polarons with a ferromagnetic component in the (001) plane which competes with an antiferromagnetic coupling along the $c$-axis, may in this case due to the so-called de Gennes double exchange, to distinguish from the Zener's double exchange,.

Above a $Mn^{4+}$ content of about 10% the two systems behaves differently. In the Na-doped series only one sample was found to be orthorhombic ($La_{0.95}Na_{0.05}MnO_3$) so the O-R transition appears very soon and the samples are always rhombohedral for $Mn^{4+}$ content greater than ~ 10%. On the opposite, for the Ca-doped samples, the orthorhombic crystal structure was found for all the samples while the rhombohedral one was observed only in connection with the highest $Mn^{4+}$ content (30%). This difference in the crystal structure can be accounted for the small, but significative, difference in the tolerance factor between the two doping series which is in turn connected with the difference between the ionic radii of the two cations.

Of course this difference has a crucial role on the physical properties of the manganites. It is well know that the optimal conditions for the polaronic hopping between charge carriers localized on the $Mn^{3+}$ ions is the presence of a Mn-O-Mn bond angle of 180° [5,31]. This is connected to the width of the narrow σ* band below $T_c$ which is given (in the tight binding approximation) by:

$$W_\sigma \approx \varepsilon_\sigma \lambda_\sigma^2 \cos\phi \langle \cos(\theta_{ij}/2) \rangle \qquad (11)$$

where $\theta_{ij}$ is the angle between spins on neighboring Mn atoms [31]. Alignment of the spins below $T_c$ favors stabilization of itinerant-electron states ($\tau_h < \omega_0^{-1}$) relative to localization of the carriers in polaronic states ($\tau_h > \omega_0^{-1}$) where $\tau_h \approx \hbar / W$ is the time for an electron to tunnel from a $Mn^{3+}$ to a $Mn^{4+}$ ion. The physical properties of manganites tend to be optimized by increasing the Mn-O-Mn bond angle and also by reducing the Mn-O bond length which is also responsible of direct effect on the σ* band width.

Parallel to the crystal structure evolution with the $Mn^{4+}$ amount also the transition temperatures in the resistivity curves ($T_\rho$) and in the magnetization plots ($T_c$) increase by increasing the oxidation of manganese array. For both system is it possible to appreciate that nearly all the points lie on common curves (see Figures 12 and 13). This is a general trend already observed in analogous systems. Our approach, anyway, allowed to tune the amount of $Mn^{4+}$ by both varying the cation doping and the oxygen content. As can be seen, the $Mn^{4+}$ amount appears to be the main variable which influences the physical properties of the considered materials but a significant role is also played by the point-defect associated with the oxygen non-stoichiometry.

For the Ca-doped materials a general worsening of the transport properties both in the semiconducting- and in the metallic-like phase is played by the point defects introduced in the lattice by the oxygen under-stoichiometry. Indeed, the appearance of shoulders at temperatures near $T_c$, for the samples showing only an activated $\rho(T)$ behavior, and of additional peaks or shoulders in the $\rho(T)$ curves of the samples with S-M transition, suggests the coexistence of regions with different transport features. The presence of different sample regions is also confirmed by the extent of the temperature range of the completion of the P-F transition and by the differences between the ZFC and FC magnetization curves. For the samples characterized by oxygen under-stoichiometry, affected by oxygen vacancies according to equilibrium of eq. 5, these features are more evident and a remarkable difference between $T_\rho$ and $T_c$ values is also present.

About the cation vacancies, for the orthorhombic $La_{0.9}Ca_{0.1}MnO_{3.054}$ sample $T_\rho$ and $T_c$ values are close each other and follow the general trend, differently from what has been observed on Na-doped manganites [41] (see later in the text).

Let's pass now to the La-Na-Mn-O system. In the orthorhombic phase a semiconducting transport mechanism was always observed for the Na-doped samples. The activation energy values (Table 1) confirm the known close correlation between the polaronic hopping and the crystal structure, particularly with the Mn-O bond lengths and Mn-O-Mn bond angle; moreover, a close correlation can be now also established with the presence of lattice defects created by the oxygen non-stoichiometry. For the $La_{0.95}Na_{0.05}MnO_{3+\delta}$ samples the lowest $E_a$ value corresponds to the stoichiometric sample (defect-free) while the introduction of anion vacancies ($\delta$= -0.01) increases $E_a$. For the over-stoichiometric sample ($\delta$= 0.078) in spite of a rhombohedral structure and a relatively high $Mn^{4+}$ amount, the charge carrier hopping is little hindered by the presence of about 2.5% cation vacancies on the manganese site. Anyway, this is the only sample of this series which shows the insurgence of a S-M transition ($T_\rho$ = 120 K) even if the high defect concentration makes the residual resistivity high. For all these samples, anyway, a magnetic P-F transition can be detected. The maximum $T_c$ value (Table 1) corresponds to the sample with $\delta$= 0 (152 K) with a $Mn^{4+}$ content of about 10 %. Let's note that for an analogous doping level, in an oxygen stoichiometric sample for the La-Ca-Mn-O system, $T_c$ is about 155 K [30]. The introduction of anion vacancies according to equilibrium (5) causes a reduction of $T_c$ to 112 K. For the over-stoichiometric sample, the relatively high amount of cation vacancies, introduced in accordance with equilibrium (2), causes, as well, a reduction of $T_c$ to 117 K. The shape of the magnetization curves again indicates the crucial role of the defects introduced by the oxygen non-stoichiometry, the greater difference between FC and ZFC curves pertaining to the under-stoichiometric sample of the $x$ = 0.05 series.

As regards the $La_{0.87}Na_{0.13}MnO_{3+\delta}$ samples, the increased extrinsic cation doping makes all of them rhombohedral. The more symmetric structure enables an easier polaron hopping which results in lower $E_a$ values with respect to the $x$ = 0.05 samples. Among the $x$ = 0.13 samples, the one without point defects ($\delta$= 0) allows the easiest carrier motion; moreover, this is also the sample with the highest $T_\rho$ (303 K) and the sharper S-M transition. The creation of anion vacancies reduces $T_\rho$ to 260 K for the sample with $\delta$= -0.03, which displays as well a sharp S-M transition, while the further reduction of oxygen content ($\delta$= -0.04) inhibits the S-M transition and leads to a broad transition with $T_\rho$ = 195 K between two semiconducting regimes. Also for all these samples a P-F transition is observed in the magnetization curves. For $\delta$= 0 and $\delta$= -0.03 $T_c$ is close to $T_\rho$ while for $\delta$= -0.04 a discrepancy is observed. For the lowest oxygen content, anyway, the greater difference in the FC-ZFC curves is detected.

Completely different is the role of cation vacancies. We characterized two samples with the same "nominal" $Mn^{4+}$ content (26%): $x$ = 0.13 and $\delta$= 0, $x$ = 0.05 and $\delta$= 0.078. The first sample, practically defects free, as told above, shows the highest $T_\rho$ and $T_c$ and the lowest resistivity values. For the other sample, the great oxygen over-stoichiometry results in a large amount of cation vacancies. The vacancies on the Mn site can interrupt the magnetic interaction paths between $Mn^{3+}$-$Mn^{4+}$ ions thus giving rise to the observed decrease of the $T_c$.

For both systems it was established that the anion vacancies seem to primarily influence the transport properties by both annihilating $Mn^{4+}$ ions and creating trapping centers and lattice distortion for the charge carriers while act on the magnetic properties only by the overall decrease on the $Mn^{4+}$ amount and by the creation of sample regions with different $T_c$ values due to different $Mn^{4+}$ concentrations and magnetic interactions. So, anomalous *ρ(T)* behaviors and differences between ZFC and FC *M(T)* curves can be ascribed to very different responses of the different sample regions as well evidenced by EPR measurements and the overall decrease of conductivity can be also due to region boundary effects.

Concerning the cation vacancies it was observed that they play a different role in the two systems; in particular they deeply affect the magnetic properties for the Na-doped samples while only marginal effects were observed in the Ca-doped materials.

The main difference between the Na and Ca-doped manganites is the crystal structure, rhombohedral for the Na-doped materials for $Mn^{4+}$ > 10% and orthorhombic for Ca-doped ones for $Mn^{4+}$ < 30%. This may suggest that the relative amount of cation vacancies on the A and B sites does not follow what proposed by [22] for the undoped material. Indeed a different energy for the defects-formation arises for the two distinct crystal structures, as evidenced by some recent computational results [25] and consequently different distribution between the two sites may be present. We have planned further experiments in order to clarify this point.

In conclusion, in this review the role of the different defects, *i.e.* oxygen vacancies and cation vacancies, related to oxygen non-stoichiometry on the structural, transport and magnetic properties of Calcium and Sodium doped manganites is discussed. The results obtained suggest taking in serious account the role of oxygen content when dealing with this kind of materials.

# Tables

TABLE 1

| Sample | *a* (Å) | *b*(Å) | *c* (Å) | *V*(Å$^3$) | $T_\rho$ (K) | $T_c$ (K) | % Mn(IV) |
|---|---|---|---|---|---|---|---|
| $La_{0.9}Ca_{0.1}MnO_{3.054}$ | 5.478 | 5.519 | 7.758 | 58.64 | 210(190) | 202 | 22 |
| $La_{0.9}Ca_{0.1}MnO_3$ | 5.513 | 5.527 | 7.7815 | 59.27 | (160) | 170 | 10 |
| $La_{0.9}Ca_{0.1}MnO_{2.986}$ | 5.515 | 5.542 | 7.7618 | 59.31 | (140) | 150 | 7.5 |
| $La_{0.7}Ca_{0.3}MnO_3$ | 5.478 | 5.478 | 13.205 | 57.23 | 271 | 270 | 30 |
| $La_{0.7}Ca_{0.3}MnO_{2.99}$ | 5.462 | 5.488 | 7.729 | 57.92 | 261(239) | 259 | 28 |
| $La_{0.7}Ca_{0.3}MnO_{2.975}$ | 5.466 | 5.492 | 7.723 | 57.96 | 229(257) | 258 | 25 |
| $La_{0.95}Na_{0.05}MnO_{3.078}$ | 5.514 | 5.514 | 13.328 | 58.49 | 120 | 117±5 | 26 |
| $La_{0.95}Na_{0.05}MnO_3$ | 5.517 | 5.534 | 7.785 | 59.43 | - | 152±5 | 10 |
| $La_{0.95}Na_{0.05}MnO_{2.99}$ | 5.525 | 5.587 | 7.753 | 59.83 | - | 112±5 | 8 |
| $La_{0.87}Na_{0.13}MnO_3$ | 5.512 | 5.512 | 13.351 | 58.54 | 303 | 305 | 26 |
| $La_{0.87}Na_{0.13}MnO_{2.97}$ | 5.533 | 5.533 | 13.366 | 59.06 | 260 | 267 | 20 |
| $La_{0.87}Na_{0.13}MnO_{2.96}$ | 5.535 | 5.535 | 13.370 | 59.12 | 195 | 138 | 18 |

TABLE 2

| Sample | $E_a$ (meV) | $\rho_0$ (Ωcm) | $\rho_2$ ($10^{-6}$ΩcmK$^{-2}$) | $\rho_{4.5}$ ($10^{-12}$ΩcmK$^{-4.5}$) |
|---|---|---|---|---|
| $La_{0.9}Ca_{0.1}MnO_{3.054}$ | 132.5 | 0.8045 | 2.24 *10-7 | 162.6 |
| $La_{0.9}Ca_{0.1}MnO_3$ | 170 | | | |
| $La_{0.9}Ca_{0.1}MnO_{2.986}$ | 217 | | | |
| $La_{0.7}Ca_{0.3}MnO_3$ | 151 | 0.043 | 2.14 | 2.6 |
| $La_{0.7}Ca_{0.3}MnO_{2.99}$ | 154 | 0.062 | 3.85 | 6.8 |
| $La_{0.7}Ca_{0.3}MnO_{2.975}$ | 190 | 0.105 | 3.88 | 21 |
| $La_{0.95}Na_{0.05}MnO_{3.078}$ | 210 | | | |
| $La_{0.95}Na_{0.05}MnO_3$ | 196 | | | |
| $La_{0.95}Na_{0.05}MnO_{2.99}$ | 235 | | | |
| $La_{0.87}Na_{0.13}MnO_3$ | 72 | 0.044 | 2.25 | 2.3 |
| $La_{0.87}Na_{0.13}MnO_{2.97}$ | 105 | 0.184 | 11.6 | 5 |
| $La_{0.87}Na_{0.13}MnO_{2.96}$ | 138 | | | |

## Tables Captions

**Table 1.** Lattice parameters (*a*, *b* and *c*), cell volumes (*V*), transition temperatures ($T_\rho$ and $T_c$) and estimated Mn(IV) content for the samples considered in the paper.
**Table 2.** Activation energies ($E_a$), according to (9), and fit parameters, according to (10), for the samples considered in the paper.

## Figure Captions

**Figure 1** – X-ray powder patterns for $La_{1-x}Ca_xMnO_{3+\delta}$ with $x$ = 0.1 (top) and $x$ = 0.3 (bottom).
**Figure 2** – Electrical resistivity ($\rho$) *vs*. temperature for $La_{0.9}Ca_{0.1}MnO_{3+\delta}$ samples with $\delta$= 0 (a), $\delta$= –0.014 (b) and $\delta$= 0.054 (c).
**Figure 3** – Electrical resistivity ($\rho$) *vs*. temperature for $La_{0.7}Ca_{0.3}MnO_{3+\delta}$ samples.
**Figure 4** – ZFC and FC molar magnetization at 100 Oe for the $La_{0.9}Ca_{0.1}MnO_{3+\delta}$ samples: $\delta$ = 0 (open squares), $\delta$= –0.014 (open triangles) and $\delta$= 0.054 (open circles).
**Figure 5** – ZFC and FC molar magnetization at 100 Oe for the $La_{0.7}Ca_{0.3}MnO_{3+\delta}$ samples: $\delta$ = 0 (open squares), $\delta$= –0.01 (open triangles) and $\delta$= -0.025 (open circles).
**Figure 6** – X-ray powder patterns for $La_{1-x}Na_xMnO_{3+\delta}$ with $x$ = 0.05 (bottom) and $x$ = 0.13 (top).
**Figure 7 -** Logarithm of electrical resistivity *vs*. temperature for $La_{0.95}Na_{0.05}MnO_{3+\delta}$ samples.
**Figure 8 -** Logarithm of electrical resistivity *vs*. temperature for $La_{0.87}Na_{0.13}MnO_{3+\delta}$ samples.
**Figure 9-** ZFC and FC molar magnetization at 100 Oe for the $La_{0.95}Na_{0.05}MnO_{3+\delta}$ samples.
**Figure 10 -** ZFC and FC molar magnetization at 100 Oe for the $La_{0.87}Na_{0.13}MnO_{3+\delta}$ samples.
**Figure 11** - Temperature behavior of the derivative EPR signals of the A and B regions for the $La_{1-x}Na_xMnO_{3+\delta}$ sample with $x$ = 0.05 and $\delta$= 0. In (b) the signal intensity is amplified by a factor 5.
**Figure 12** - Transition temperatures ($T_\rho$ and $T_c$) *vs*. $Mn^{4+}$ content for the $La_{1-x}Ca_xMnO_{3+\delta}$ system; O and R refer to the orthorhombic and rhombohedral crystal structure, respectively; SP means semiconducting-paramagnet; SF semiconducting-ferromagnetic and MF metallic-ferromagnetic.
**Figure 13** - Transition temperatures ($T_\rho$ and $T_c$) *vs*. $Mn^{4+}$ content for the $La_{1-x}Na_xMnO_{3+\delta}$ system; O and R refer to the orthorhombic and rhombohedral crystal structure, respectively; SP means semiconducting-paramagnet; SF semiconducting-ferromagnet and MF metallic-ferromagnet.

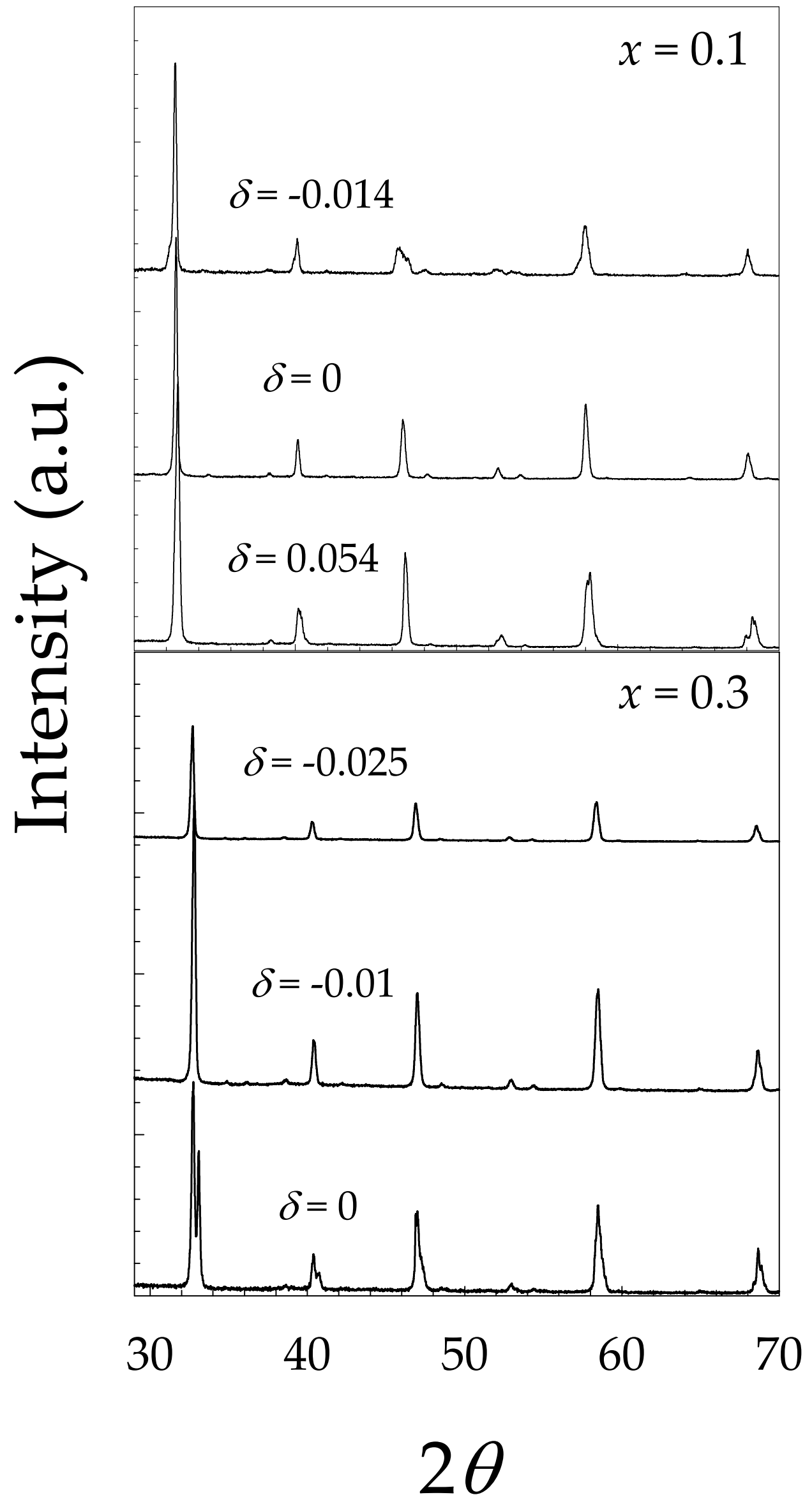

Figure 1

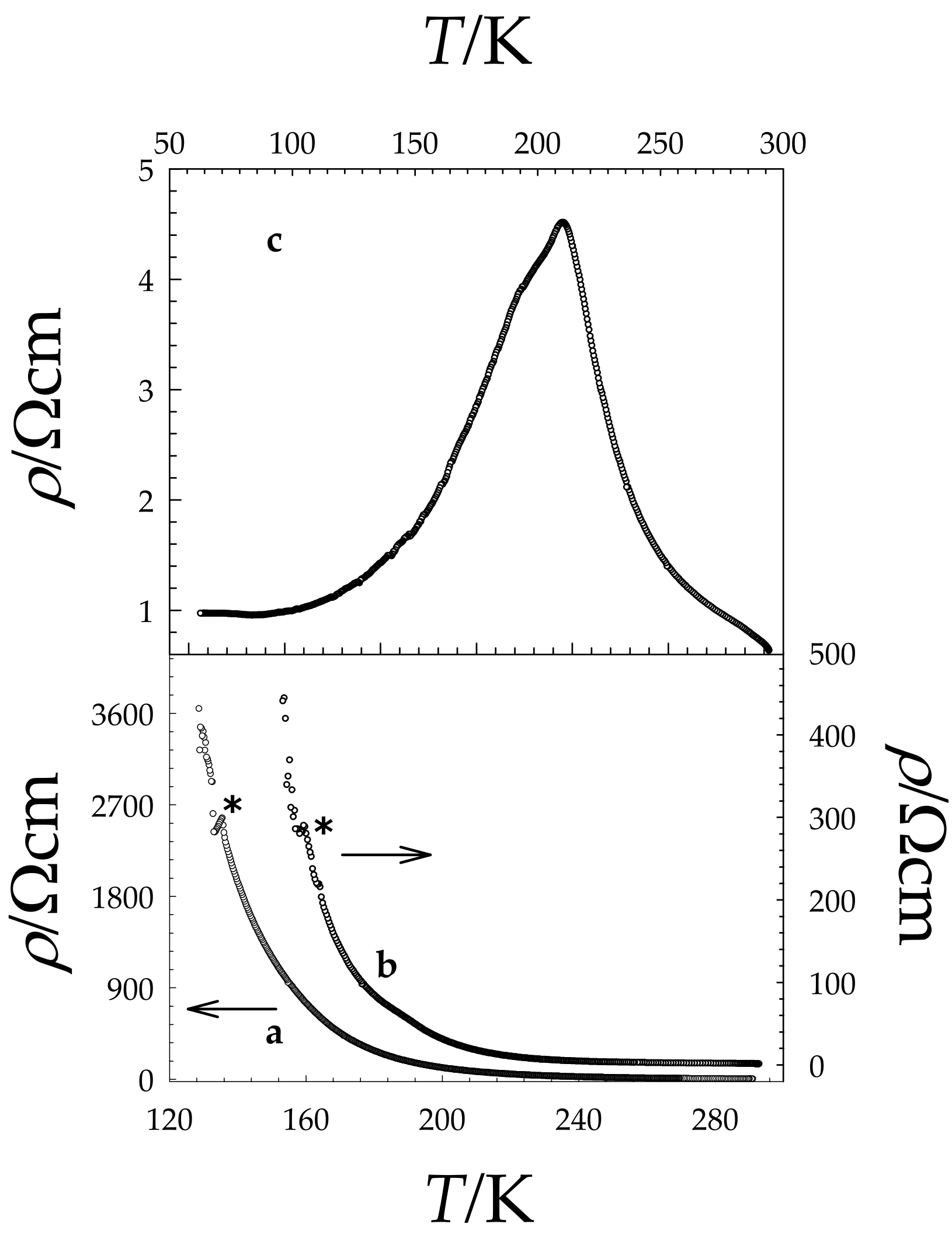

Figure 2

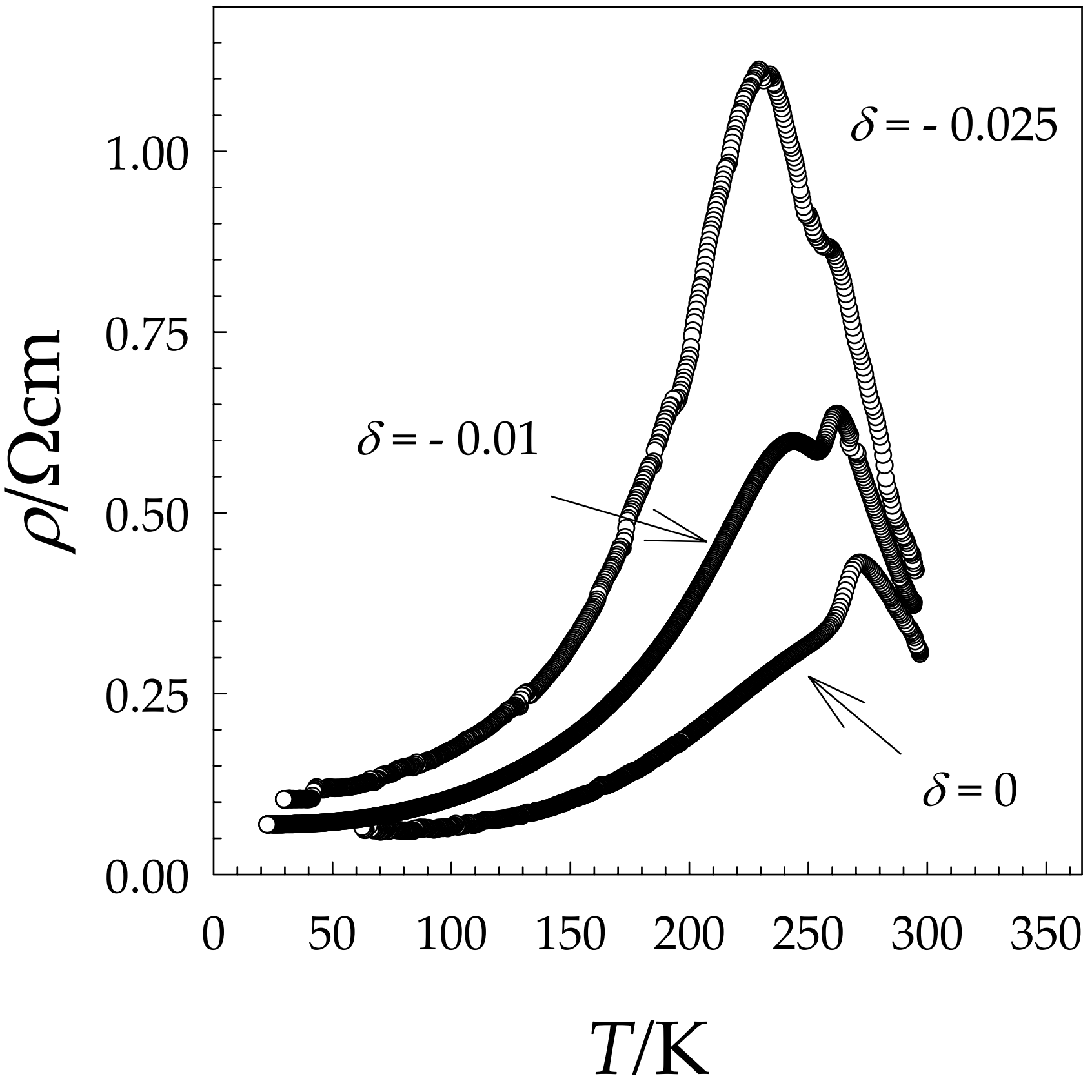

Figure 3

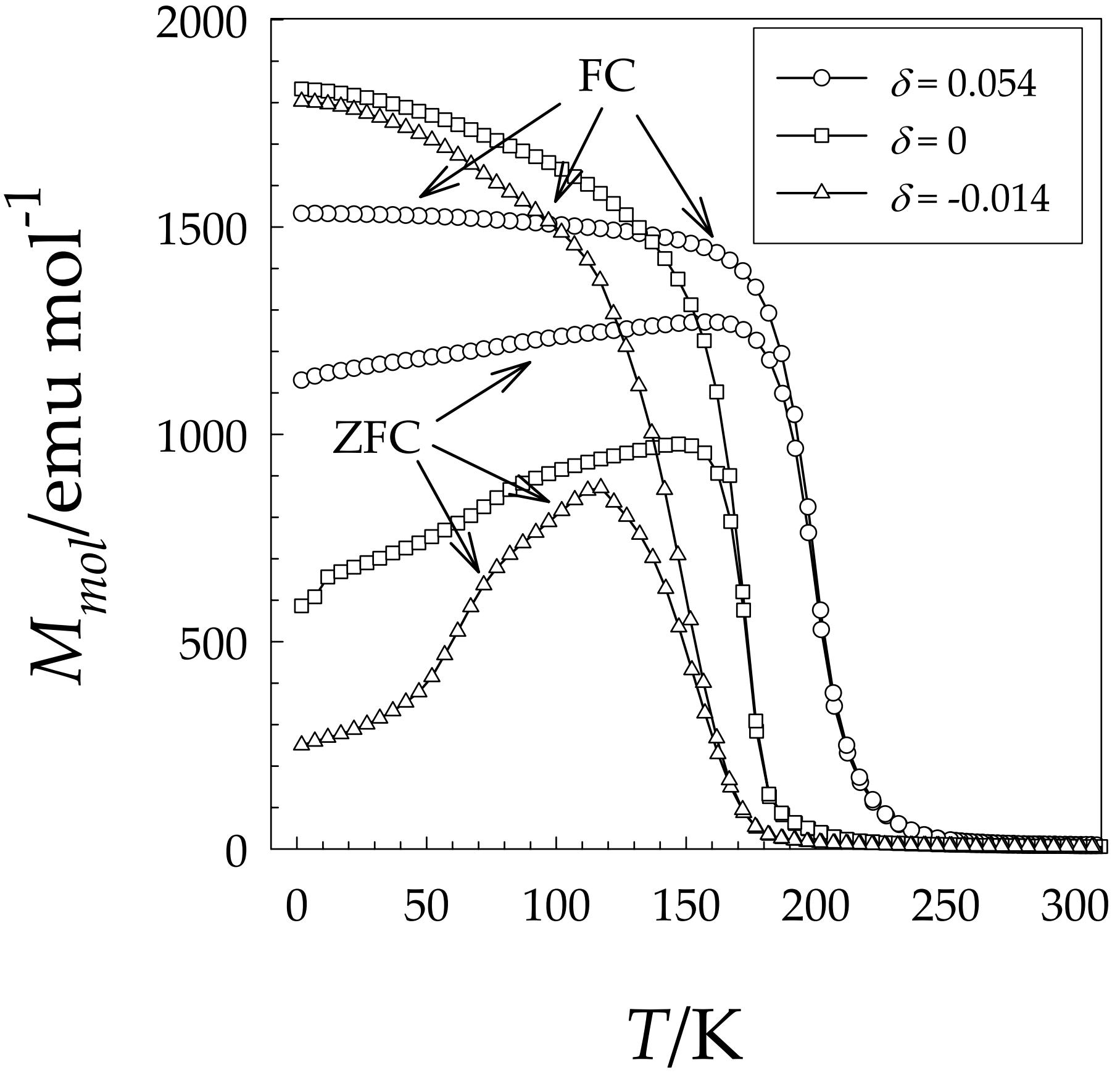

Figure 4

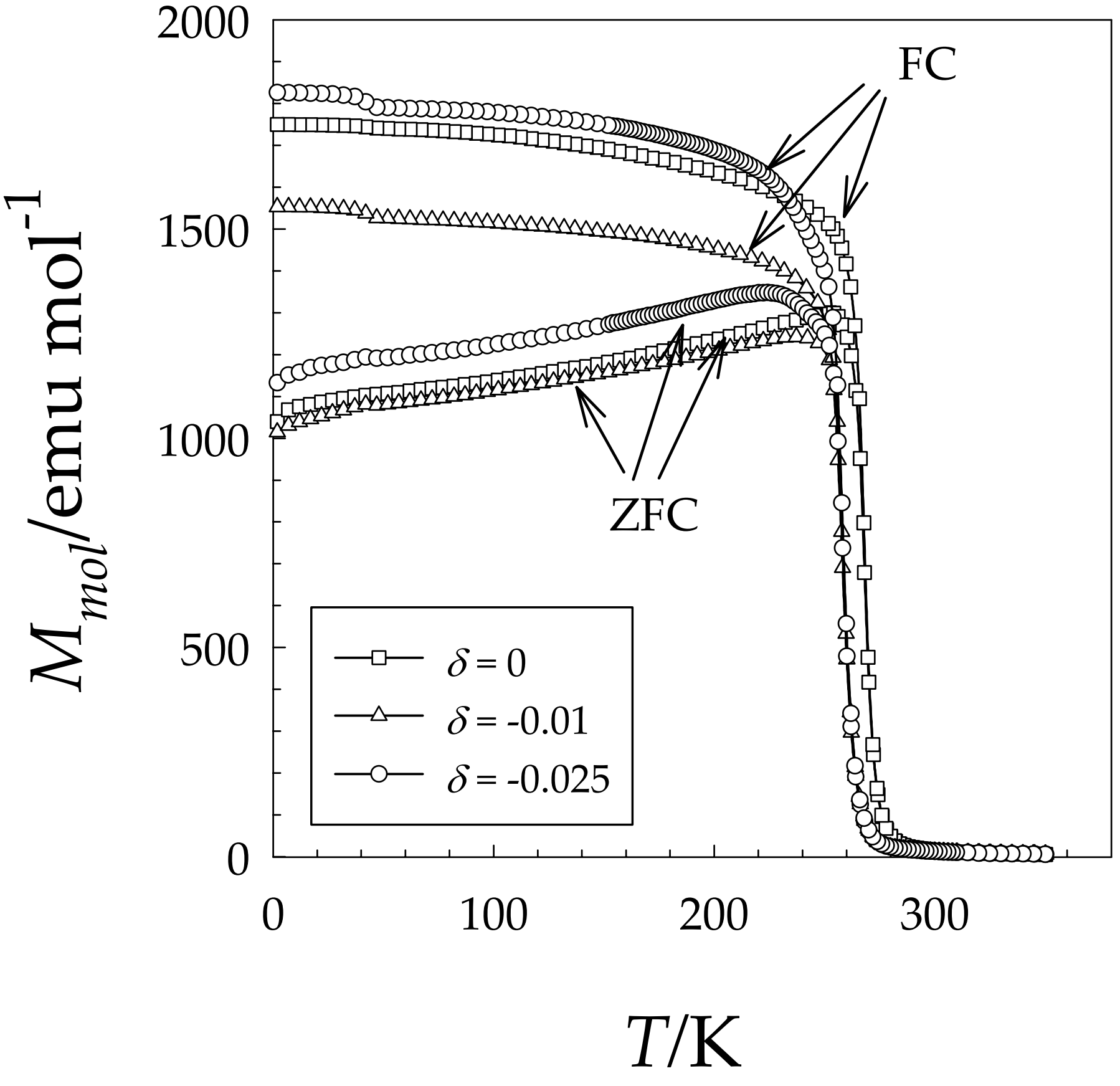

Figure 5

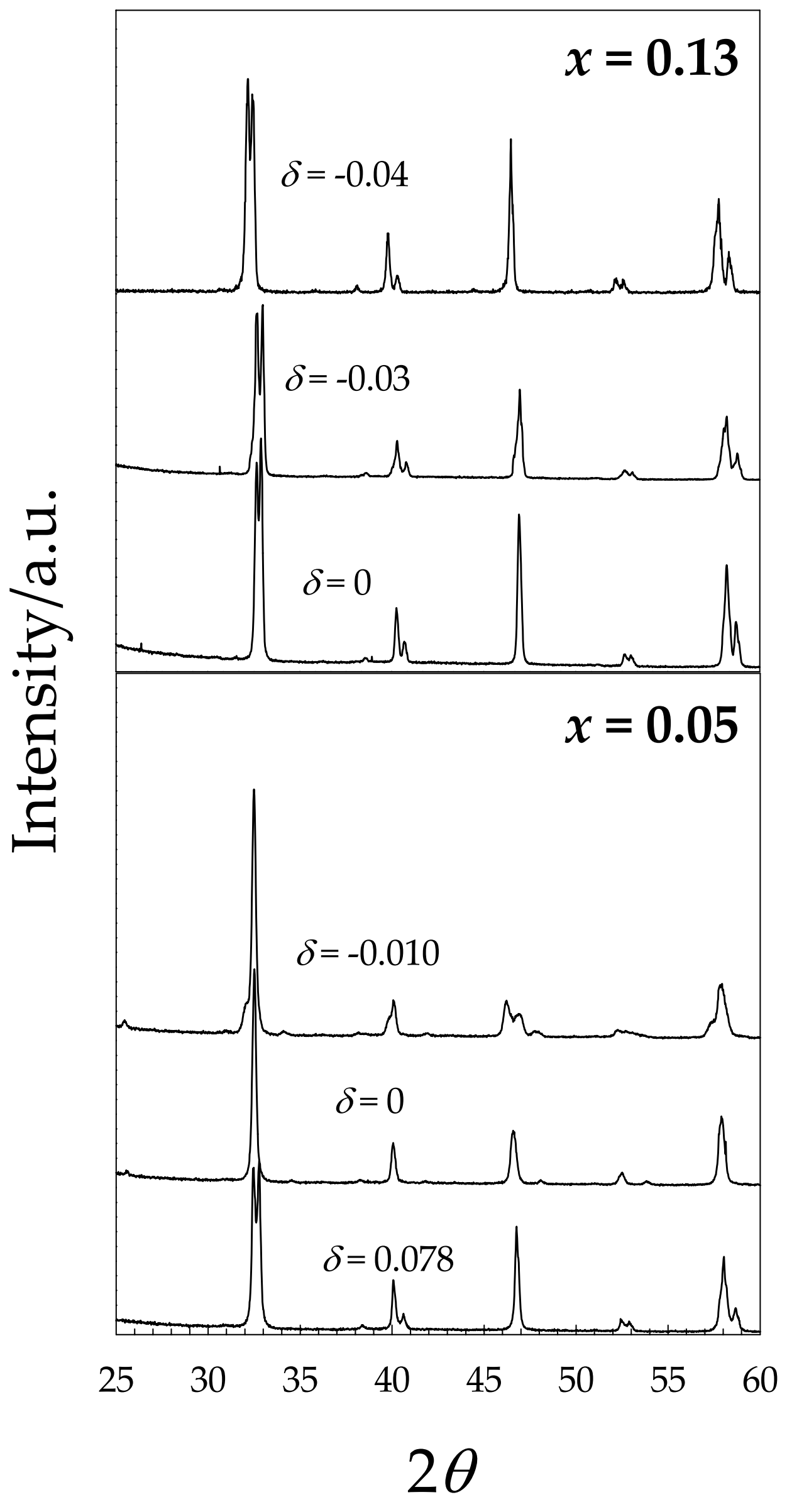

Figure 6

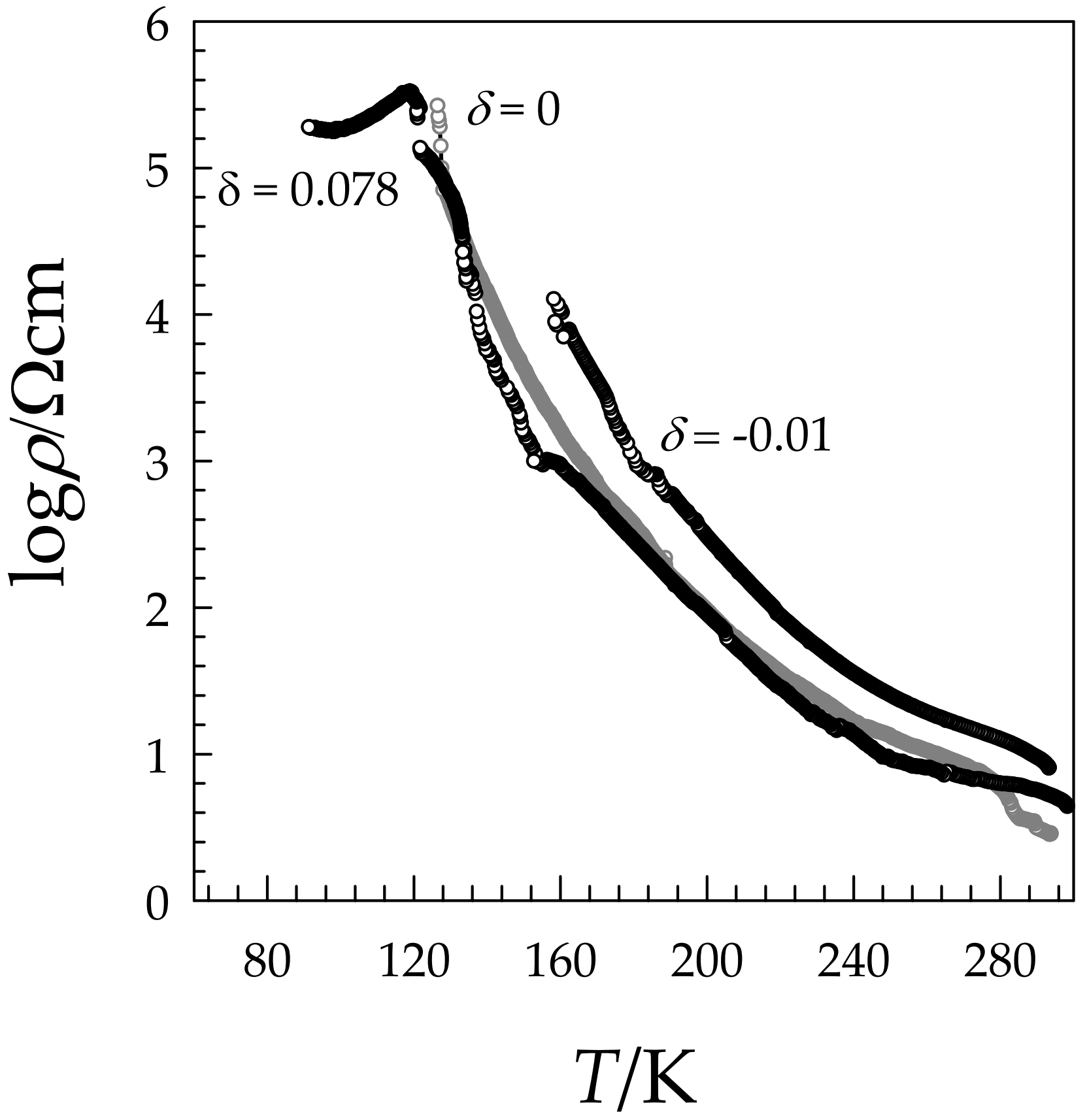

Figure 7

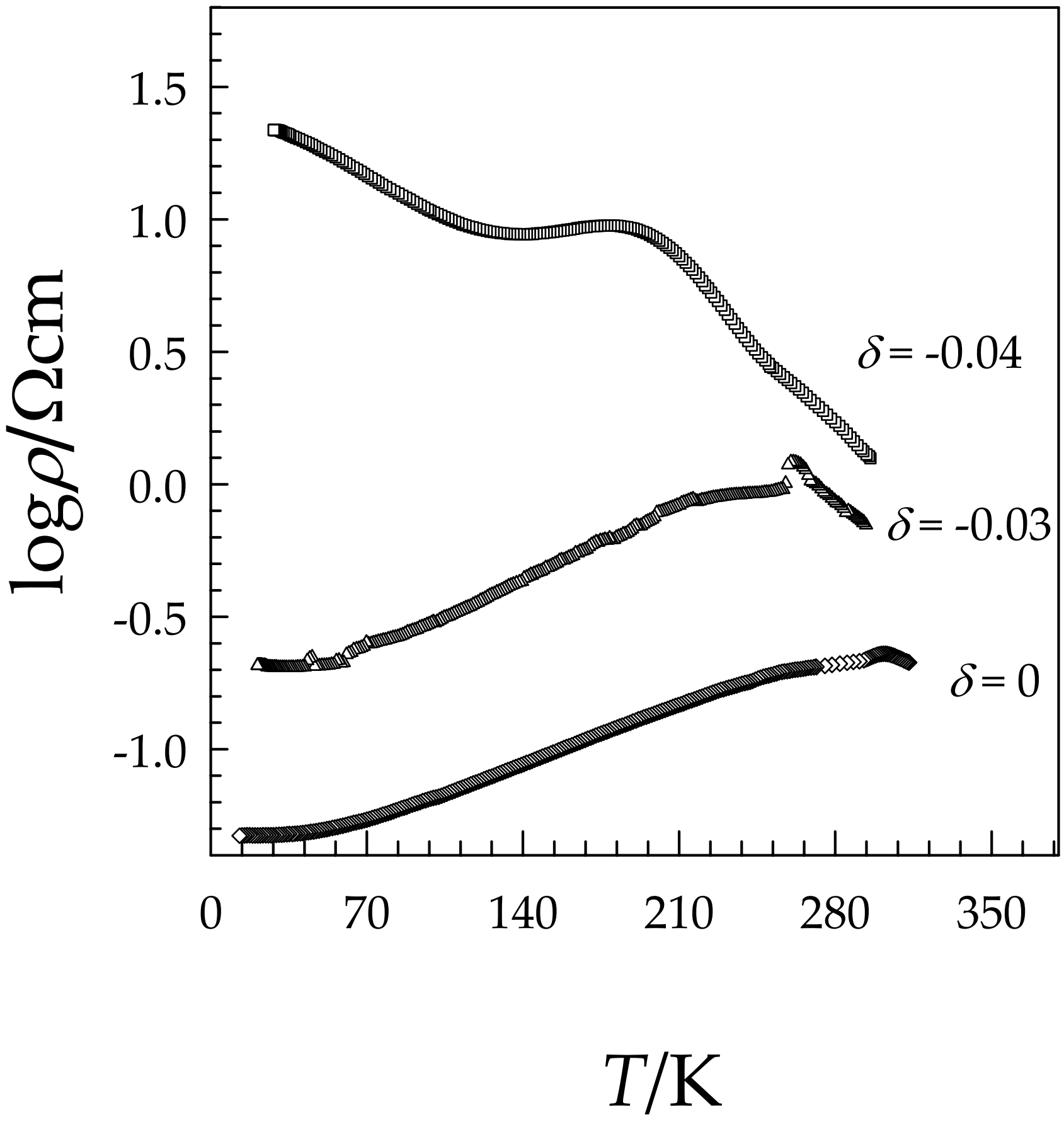

Figure 8

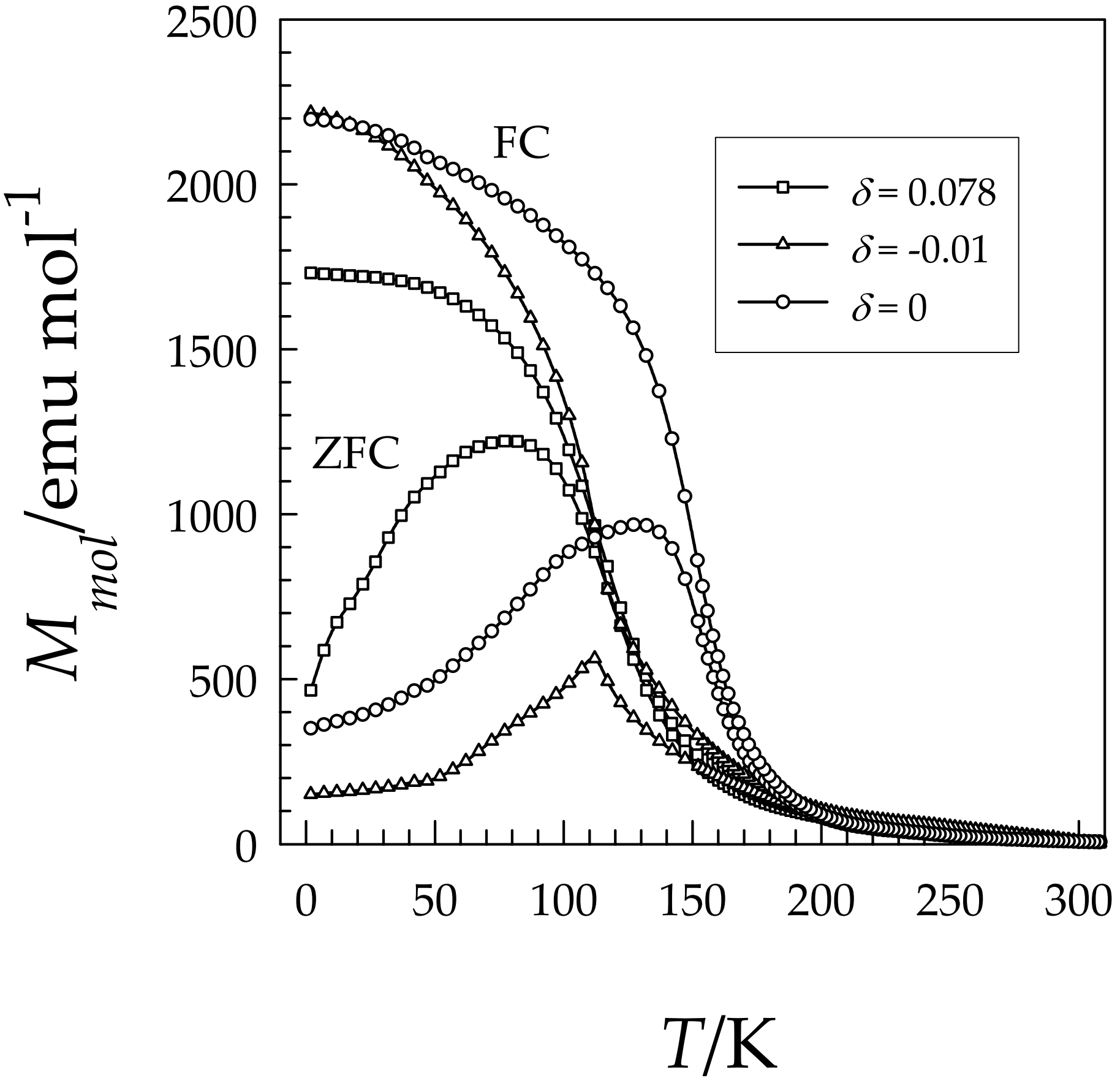

Figure 9

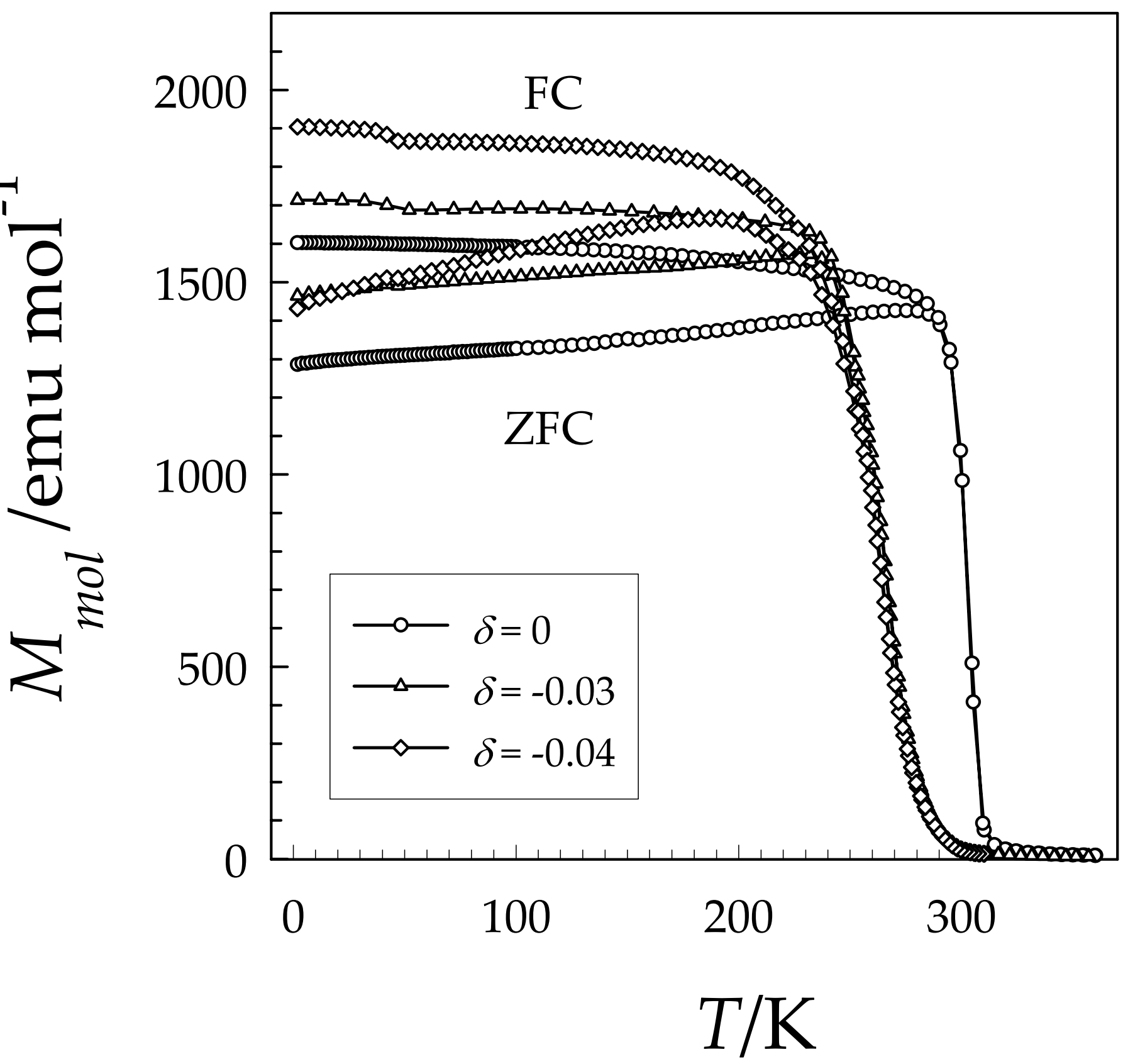

Figure 10

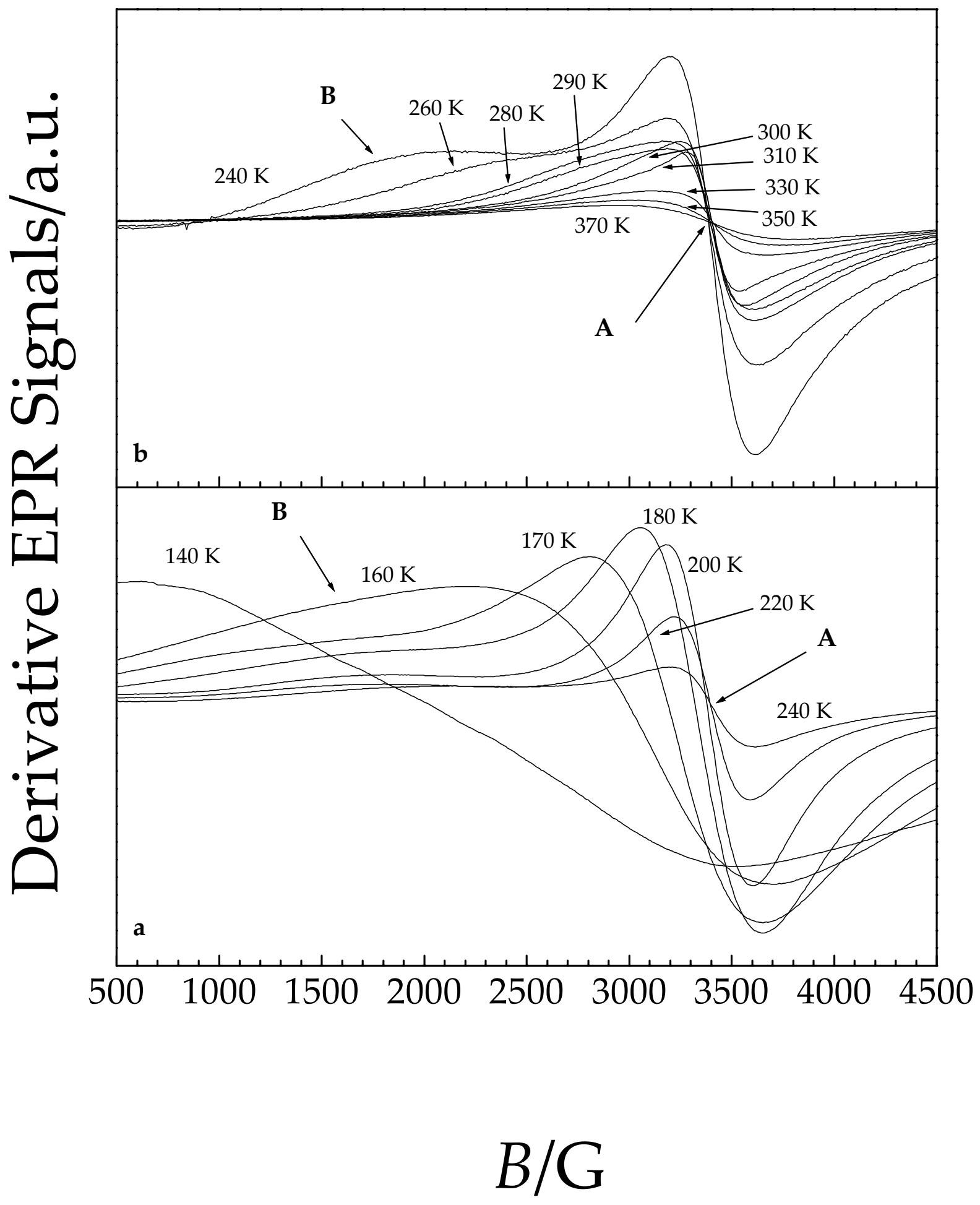

Figure 11

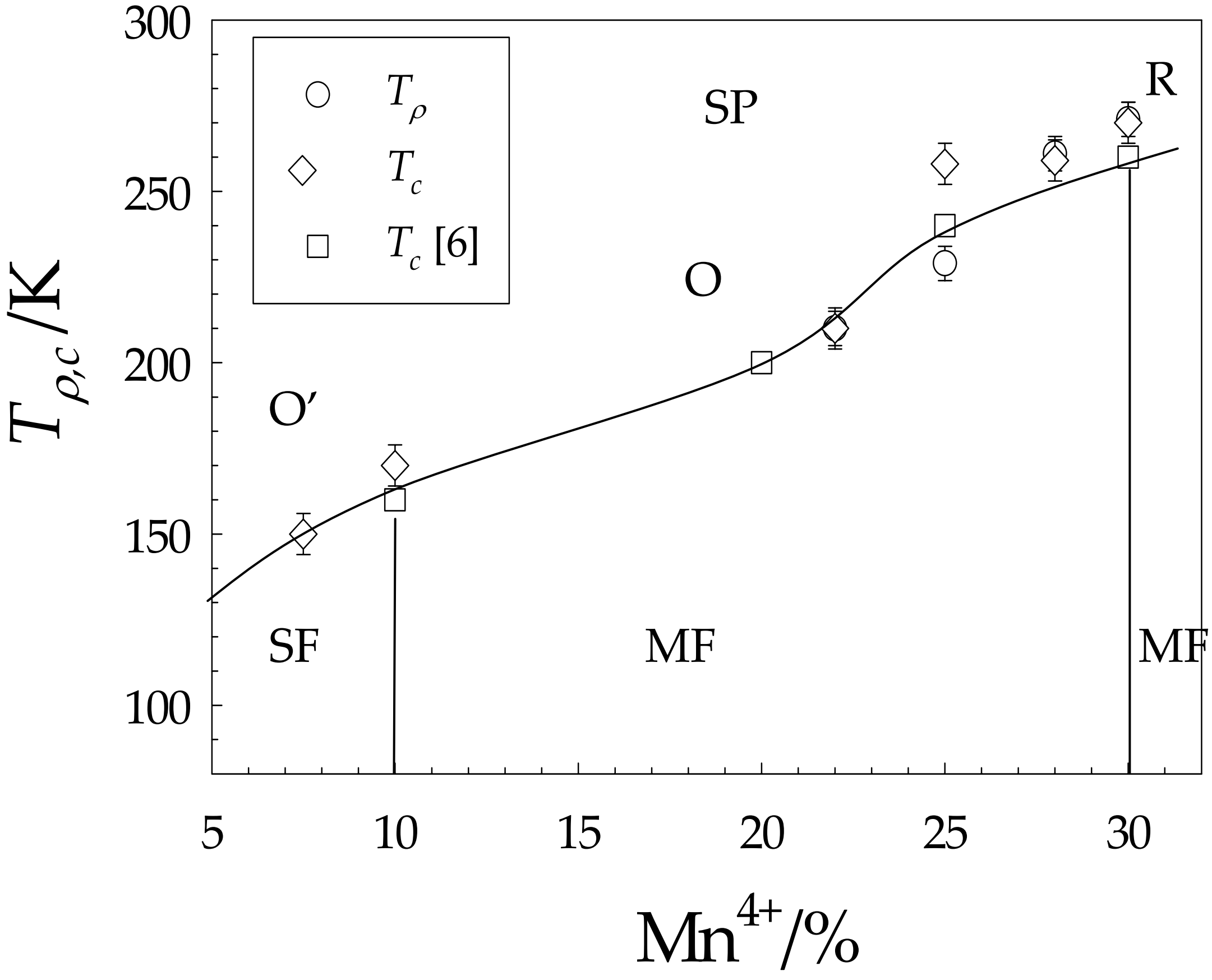

Figure 12

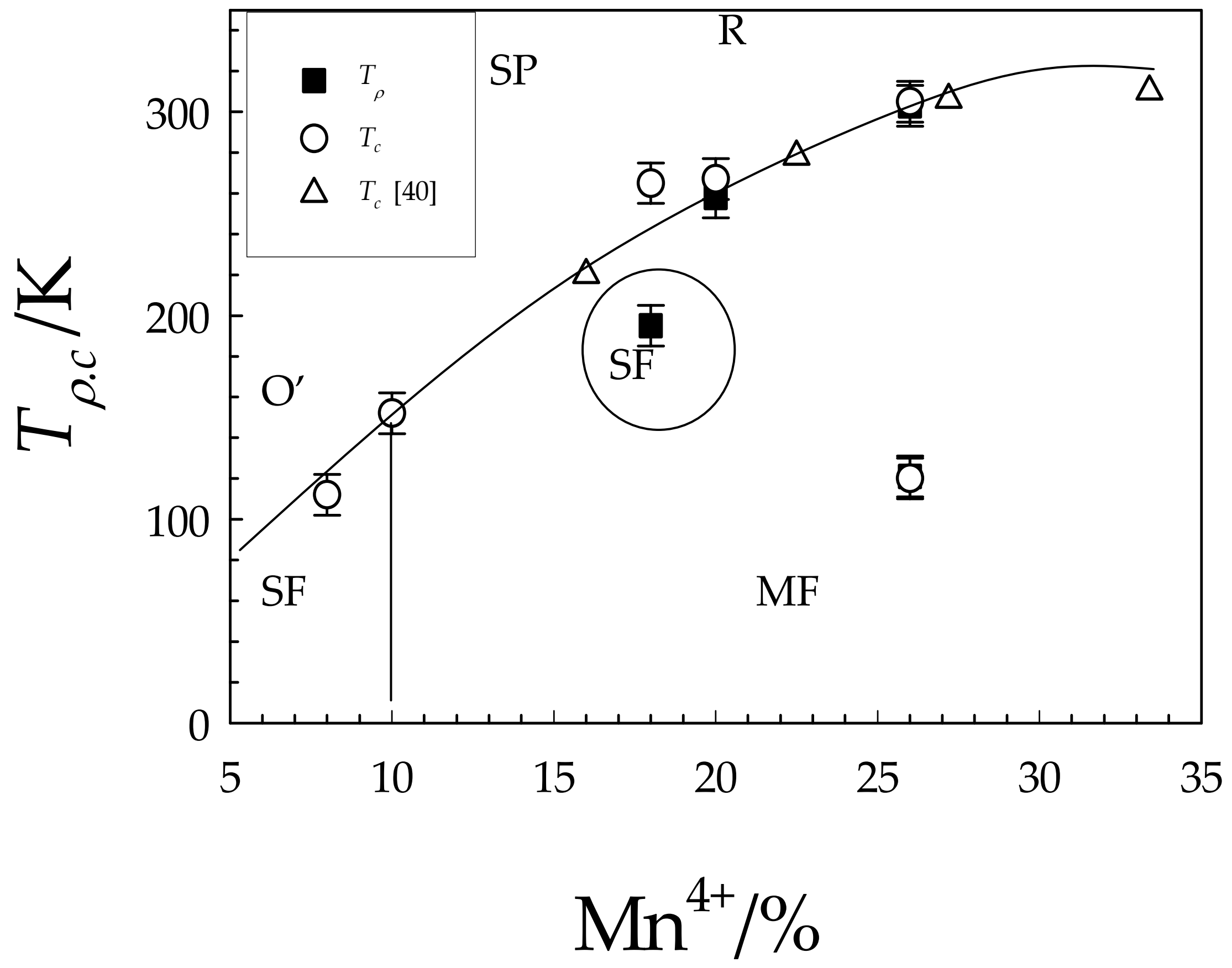

Figure 13